\documentclass[aps,twocolumn,prb,showpacs,reprint,amsmath,amssymb,superscriptaddress]{revtex4-1}

\usepackage{graphicx}
\usepackage{dcolumn}
\usepackage{bm}
\usepackage{array}

\begin{document}

\title{Determining the stability and activation energy of Si acceptors in AlGaAs using quantum interference in an open hole quantum dot.}

\author{D.J.~Carrad}
\affiliation{School of Physics, University of New South Wales,
Sydney NSW 2052, Australia}

\author{A.M.~Burke}
\affiliation{School of Physics, University of New South Wales,
Sydney NSW 2052, Australia}

\author{O.~Klochan}
\affiliation{School of Physics, University of New South Wales,
Sydney NSW 2052, Australia}

\author{A.M.~See}
\affiliation{School of Physics, University of New South Wales,
Sydney NSW 2052, Australia}

\author{A.R.~Hamilton}
\affiliation{School of Physics, University of New South Wales,
Sydney NSW 2052, Australia} 

\author{A.~Rai}
\affiliation{Angewandte Festk\"{o}rperphysik, Ruhr-Universit\"{a}t
Bochum, D-44780 Bochum, Germany}

\author{D.~ Reuter}
\affiliation{Angewandte Festk\"{o}rperphysik, Ruhr-Universit\"{a}t
Bochum, D-44780 Bochum, Germany}

\author{A.D.~Wieck}
\affiliation{Angewandte Festk\"{o}rperphysik, Ruhr-Universit\"{a}t
Bochum, D-44780 Bochum, Germany}

\author{A.P.~Micolich}
\email{adam.micolich@nanoelectronics.physics.unsw.edu.au}
\affiliation{School of Physics, University of New South Wales,
Sydney NSW 2052, Australia}

\date{\today}

\begin{abstract}
We fabricated an etched hole quantum dot in a Si-doped (311)A
AlGaAs/GaAs heterostructure to study disorder effects via
magnetoconductance fluctuations (MCF) at millikelvin temperatures.
Recent experiments in electron quantum dots have shown that the MCF
is sensitive to the disorder potential created by remote ionised
impurities. We utilize this to study the temporal/thermal stability
of Si acceptors in $p$-type AlGaAs/GaAs heterostructures. In
particular, we use a surface gate to cause charge migration between
Si acceptor sites at $T~=~40$~mK, and detect the ensuing changes in
the disorder potential using the MCF. We show that Si acceptors are
metastable at $T~=~40$~mK and that raising the device to a
temperature $T~=~4.2$~K and returning to $T~=~40$~mK is sufficient
to produce complete decorrelation of the MCF. The same decorrelation
occurs at $T \sim 165$~K for electron quantum dots; by comparing
with the known trap energy for Si DX centers, we estimate that the
shallow acceptor traps in our heterostructures have an activation
energy $E_{A} \sim 3$~meV. Our method can be used to study charge
noise and dopant stability towards optimization of semiconductor
materials and devices.
\end{abstract}

\maketitle

\section{Introduction}
A significant outcome of device miniaturization is the semiconductor
quantum dot -- a device where electrons or holes are confined in all
three spatial dimensions at a length scale comparable to the
electron or hole wavelength. These can be made in any semiconductor,
in principle, but the $n$-type AlGaAs/GaAs heterostructure is
predominant for electrical studies because of the high electron
mobilities that can be obtained through growth by molecular beam
epitaxy (MBE).~\cite{ChoAPL71, DingleAPL78} Small AlGaAs/GaAs dots
isolated by tunnel barriers from source and drain electron
reservoirs were used as `artificial atoms'~\cite{KastnerPT93} for
fundamental studies of the physics of few-electron
systems,~\cite{KouwenhovenRPP01} and now play a vital role in the
study of electron spin physics~\cite{HansonRMP07} and the early
development of quantum computation architectures.~\cite{LaddNat10}
Larger AlGaAs/GaAs dots with stronger coupling to the reservoirs
have been central to studies of mesoscopic conduction, quantum
interference phenomena and electron transport in the quasi-ballistic
and ballistic limits, where an electron's mean-free path is
comparable to or exceeds the dot's confinement
width.~\cite{LinJPCM02, BirdRPP03, MicolichFdP13}

A hallmark of these larger `open' dots is the appearance of
fluctuations in the magnetoconductance $G(B)$ at low temperature $T
< 1$~K, which arise from the Aharonov-Bohm
effect.~\cite{AharonovPR59, MarcusPRL92} These fluctuations are
reproducible for a given dot on a single measurement
cool-down,~\cite{MarcusPRL92} providing a `magnetofingerprint' of
the electron trajectories within the dot.~\cite{FengPRL86} Recent
experiments~\cite{SeePRL12, ScannellPRB12} highlight the important
role that the dot geometry, {\it and} the remote ionized dopant
potential it contains, play in determining these trajectories, even
in the ballistic limit. For a single electron dot with fixed
geometry, the magnetoconductance fluctuations (MCF) change markedly
upon raising the temperature to $T \sim 165$~K briefly and returning
to low $T$ to remeasure $G(B)$ for a modulation-doped AlGaAs/GaAs
heterostructure.~\cite{ScannellPRB12} However, for an undoped dot,
the MCF remain identical under thermal cycling to temperatures as
high as $300$~K.~\cite{SeePRL12} The change in MCF of the
modulation-doped electron dot arises from the spontaneous
re-organization of trapped charge in the Si doping layer at $T
\geqslant 165$~K, which alters the dot's potential landscape. The
sensitivity of the MCF to small-angle Coulomb scattering by remote
ionized dopants provides a potential new application for
semiconductor quantum dots as sensitive detectors of fluctuations in
dopant ionization state.~\cite{SeePRL12} This could be combined with
other methods if needed, e.g., switching noise measurements in
quantum point contacts (QPCs),~\cite{BuizertPRL08} to enable studies
of how dopants influence the electronic properties of a given
device, as we do here, or for optimizing materials growth to obtain
devices with high stability and low operating
noise.~\cite{BuizertPRL08, Pioro-LadrierePRB05}

Here we demonstrate the potential for investigating the temporal and
thermal stability of Si acceptors in a (311)A-oriented Si-doped
AlGaAs/GaAs heterostructure using measurements of the MCF in a hole
open quantum dot etched into the heterostructure. We do so by
utilizing the measured MCF as a magnetofingerprint of the disorder
potential generated by a given spatial charge configuration in the
partially-ionized acceptor layer. (311)A-oriented Si-doped
AlGaAs/GaAs heterostructures are a key materials platform for
studying the fundamental physics of low-dimensional hole systems,
where the much stronger spin-orbit interaction obtained via the GaAs
valence band's $p$-orbital-like nature gives rise to interesting
topological phase~\cite{YauPRL02} and spin anisotropy
~\cite{WinklerPRL00, DanneauPRL06, ChenNJP10, KlochanPRL11,
SrinivasanNL13} effects.

A significant problem with (311)A-oriented Si-doped AlGaAs/GaAs
heterostructures is the hysteresis/instability that arises when
using a surface metal `gate' electrode to electrostatically alter
the density of the two-dimensional hole gas (2DHG) formed at a
buried AlGaAs/GaAs interface within the
heterostructure.~\cite{ZailerPRB94, DaneshvarPRB97, RokhinsonSM02}
We recently reported that this hysteresis contains two contributions
at different time/energy scales: a) surface-state trapping at the
gate-heterostructure interface, and b) fluctuations in ionization
state of Si dopants.~\cite{BurkePRB12, CarradJPCM13} We suggested
acceptors dominate the gate hysteresis at $T < 4$~K by comparison
with higher temperature data in $n$-type (100)-oriented Si-doped
AlGaAs/GaAs heterostructures where the surface-state contribution to
gate hysteresis is negligible and donor fluctuations cause gate
hysteresis.~\cite{BurkePRB12} The hole quantum dot data we present
here confirms the low $T$ behavior observed for $p$-type devices
in Ref.~\cite{BurkePRB12}. We show that some fraction of the Si acceptors
in Al$_{0.34}$Ga$_{0.66}$As act as very shallow trap sites that
remain metastable, even at $T = 40$~mK. This is in stark contrast to
Si donors in Al$_{0.34}$Ga$_{0.66}$As where the activation
temperature is much greater due to the formation of DX
centers,~\cite{MooneyJAP90, BuksSST94} which are donors that can
also act as deep, long-lived traps. This capacity allows DX centers
to trap charge released by the much shallower ordinary dopants, and
gives nanoscale electron quantum devices their high stability at low
$T$.

\section{Methods}
\begin{figure}
\includegraphics[width=8cm]{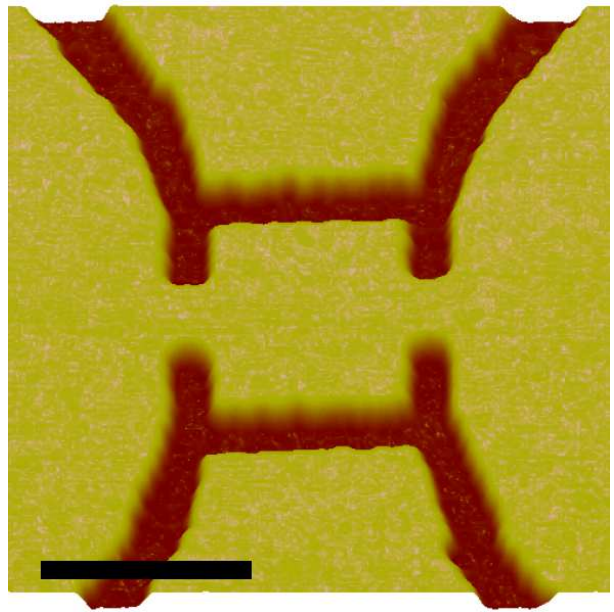}
\caption{(Color online) Atomic force micrograph of the etched quantum dot, prior to
polyimide and gate deposition. The $120$~nm deep trenches (red/dark)
define a nominally $1~\mu$m$~\times~1\mu$m square quantum dot
connected to 2DHG source and drain reservoirs via two quantum point
contacts (QPCs). The scale bar represents $1~\mu$m.}\label{fig1}
\end{figure}
The hole quantum dot was fabricated from a $p$-type modulation doped Al$_{0.34}$Ga$_{0.66}$As/GaAs heterostructure grown on half of a 2" GaAs (311)A semi-insulating substrate by molecular beam epitaxy (MBE). First, a $50$~nm undoped GaAs layer was grown, followed by an undoped, 20-period superlattice with a period of $2$~nm GaAs + $2$~nm AlAs to trap eventual segregating impurities at its interfaces. The active region was then grown, consisting of $650$~nm undoped GaAs, $35$~nm undoped Al$_{0.34}$Ga$_{0.66}$As spacer, a homogeneously Si-doped $80$~nm Al$_{0.34}$Ga$_{0.66}$As layer and a $5$~nm undoped GaAs cap to prevent oxidation of the Al-content. The Si-doped layer ($N_{Si}$ = $3.5 \times 10^{16}$~cm$^{-3}$) provides charge carriers to a high-mobility 2DHG located on the GaAs side of the interface between the undoped Al$_{0.34}$Ga$_{0.66}$As layer and the $650$~nm GaAs layer. Si was used due to its low diffusion rate compared to Be, and its amphoteric nature, which yields predominantly $n$-type incorporation on the (100)-surface, and $p$-type incorporation on the (311)A-surface.\cite{WangAPL85} At room temperature in the dark, we measured a hole density $p$ = $2.22 \times 10^{12}$~cm$^{-2}$ and mobility $\mu = 141$~cm$^2$V$^{-1}$s$^{-1}$. At $T$ = $4.2$~K we obtained $p$ = $1.63 \times 10^{11}$~cm$^{-2}$ and $\mu = 1.03 \times 10^{6}$~cm$^2$V$^{-1}$s$^{-1}$, corresponding to a large-angle scattering length $\ell = 6.8~\mu$m. Irradiation with bandgap radiation shows no persistent photoeffect, but lowers the hole density and mobility.

To monitor both the state of our MBE and the doping density in the $p$-type heterostructure, an $n$-type Al$_{0.34}$Ga$_{0.66}$As/GaAs heterostructure was grown on a second half of a semi-insulating 2" GaAs (100)-oriented wafer during the same MBE-growth. We obtained an electron density $n$ = $2.8 \times 10^{11}$~cm$^{-2}$ and mobility $\mu = 7343$~cm$^2$V$^{-1}$s$^{-1}$ at room temperature. In dividing this electron density $n$ by the thickness of the Si-doped Al$_{0.34}$Ga$_{0.66}$As layer of $80$~nm, we obtain the Si-dopant density $N_{Si} = 3.5 \times 10^{16}$~cm$^{-3}$ mentioned above, assuming a full activation of the donors and transfer of all electrons into the heterointerface at room temperature. At $T$ = $4.2$~K, we measured an electron density $n = 1.87 \times 10^{11}$~cm$^{-2}$ and mobility $\mu = 1.42 \times 10^6$~cm$^2$V$^{-1}$s$^{-1}$ before illumination and $n = 5.22 \times 10^{11}$~cm$^{-2}$ with $\mu = 2.19 \times 10^6$~cm$^2$V$^{-1}$s$^{-1}$ after illumination with bandgap radiation (persistent photo effect). This wafer was also used to verify the expected electrical performance of devices fabricated from this material growth (see Ref.\cite{BurkePRB12}).

To fabricate the hole quantum dot, a $140$~nm high Hall bar mesa was
patterned using standard photolithography and wet etching
techniques. Photolithographically defined ohmic contacts were formed
by thermally evaporating $150$~nm of AuBe alloy and annealing at
$490^\circ$C for $90$~s. The $1~\mu$m$^2$ quantum dot was defined by
electron beam lithography and wet etching with $1$:$8$:$259$
H$_2$SO$_4$:H$_2$O$_2$:H$_2$O solution to a trench depth of
$120$~nm. Figure~\ref{fig1} shows an atomic force micrograph of the
resulting dot structure. The dot was covered with a $140$~nm thick
polyimide insulating layer and an evaporation-deposited $20$~nm
Ti/$80$~nm Au top gate. A top gate voltage $V_{g}$ modulates the
hole density $p$ in the dot and surrounding 2DHG, and also alters
the dot's remote ionized dopant potential, as discussed below. The
dot's electrical conductance $G$ was measured using standard
four-terminal lock-in techniques with a $100~\mu$V excitation
voltage at $13$~Hz. The current $I_{sd}$ at $B = 0$~T ranged between
1 - 15~nA depending on the value of $V_{g}$. A Keithley $2400$ was used to control $V_{g}$
enabling continuous gate leakage current monitoring; this current
was $< 0.5$~nA throughout the experiment. The device was mounted on
a cold finger thermally linked to the mixing chamber of an Oxford
Instruments Kelvinox K$100$ dilution refrigerator enabling
measurements in the temperature range $40$~mK$~<~T~< 0.9$~K, with
two additional points at discrete temperatures of $4.2$~K and
$300$~K. The device was located inside a superconducting solenoid,
enabling a variable magnetic field $|B| < 10$~T to be applied
perpendicular to the 2DHG plane.

Our analysis relies on subtractive methods to compare sets of MCF traces. Thus we were careful to ensure each MCF trace was obtained under experimental conditions that were as identical as possible. This included ensuring that all MCF traces were obtained at fixed excitation voltage with the mixing chamber temperature $T = 40$~mK, and that MCF traces were only compared if they have the same $G$($B = 0$) to within $0.05 \times$~$2$e$^2/$h since the dot's hole density also influences the MCF. Additionally, we deliberately maintain a relatively high excitation voltage to maximise the signal-to-noise ratio of our MCF traces. Together these allow us to attribute changes in MCF solely to changes in acceptor configuration.

\begin{figure}
\includegraphics[width=8cm]{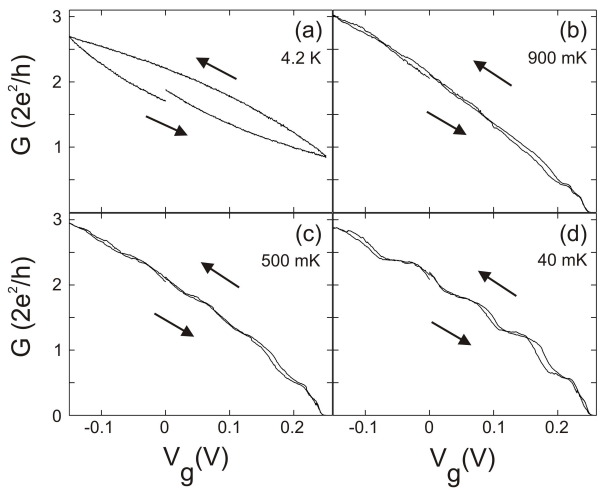}
\caption{Conductance $G$ at $B = 0$ vs gate voltage $V_{g}$ at (a)
$T = 4.2$~K (b) $900$~mK (c) $500$~mK and (d) $40$~mK at $V_{g}$
sweep rate of $1$~mV/s. The hysteresis reduces with $T$ as charge
migration between acceptor layer trap sites gradually freezes out.
Conductance plateaus also emerge with reduced $T$, generated by the
dot's entrance and exit QPCs.}\label{fig2}
\end{figure}

\section{Results}
\subsection{Temperature dependence of gate characteristics}
\label{sec:gates} Figure~\ref{fig2} shows how the conductance at
zero magnetic field $G(B~=~0)$ versus $V_{g}$ evolves as $T$ is
reduced. Significant hysteresis is observed at $T = 4.2$~K as
$V_{g}$ is first swept to $V_{g} > 0$ (depletion) and then $V_{g} <
0$ (accumulation). The anticlockwise hysteresis loop is consistent
with Ref.~\cite{BurkePRB12}, indicating transport is influenced by
gate-induced charge transfer between trap sites located between the
gate and 2DHG. The hysteresis reduces as $T$ is reduced
(Figs.~2(b-d)) due to a reduction in the available thermal energy
$kT$ relative to the trap energy $E_{A}$. Conductance quantization
develops with reduced $T$; the plateaus sit at non-integer multiples
of $G_{0} = 2e^{2}/h$ due to the non-additivity of QPCs in series
separated by less than $\ell$.~\cite{BeenakkerPRB89,
KouwenhovenPRB89}

A noteworthy aspect is that the charge migration/hysteresis means
our devices' electrical characteristics evolve with both time $t$
and experimental parameter history. Thus Fig.~2 should be
considered only as a `snapshot' of the device at that time. In
particular, readers should be cautious in using Fig.~2(d) to map
$G$ to $V_{g}$ for Figures~3-5, as the $G(B~=~0)$ versus $V_{g}$
characteristic offsets horizontally depending on device parameter
history. To deal with this, we returned to particular $G(B~=~0)$
values where quantized conductance plateaus occur to compare MCF,
rather than comparing MCF at a specific $V_{g}$. Simply returning to
a given $V_{g}$ results in much larger changes in $G(B~=~0)$ and the
MCF than in Figures~3-5.

\begin{figure*}
\includegraphics[width=15cm]{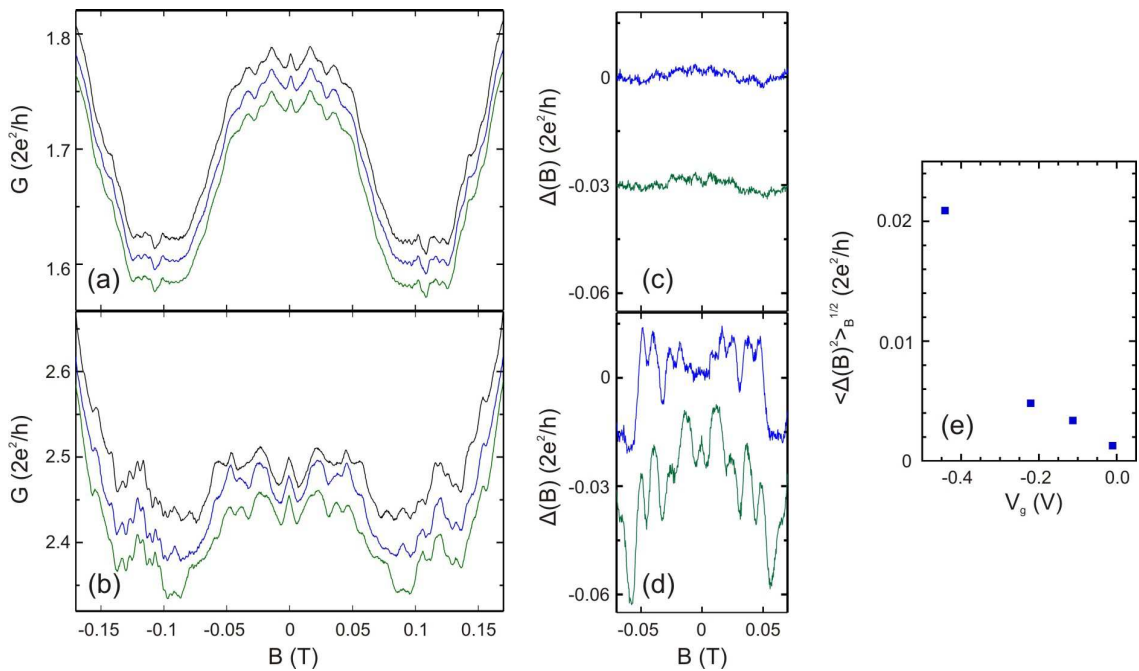}
\caption{(Color online) Magnetoconductance $G(B)$ with $G(B~=~0)$ set to the
plateaus at (a) $1.8$ and (b) $2.5~G_{0}$, using $V_{g} = -58.5$ and
$-156$~mV, respectively. In each case three traces are obtained at
times $t = 0$ (black), $8$ (blue) and $15$ hours (green) after
setting $V_{g}$. Traces offset vertically for clarity by (from top)
$0$, $0$ and $-0.02~G_{0}$ in (a) and $0$, $-0.04$ and $-0.07~G_{0}$
in (b). The background $G$ suffers a small drift between traces
(accounted for in vertical offsets) due to changes in trapped charge
in the 2DHG source and drain reservoirs (see text). (c) and (d) show
the corresponding conductance difference $\Delta(B) = \delta G(B) -
\langle \delta G(B) \rangle_{B}$ between the $t = 8$ and $0$ hour
traces (blue) and $t = 15$ and $0$ hour traces (green) vs $B$. Each
green trace is offset by $-0.03~G_{0}$ (e) root-mean-square
conductance difference $\langle \Delta (B)^{2}
\rangle_{B}^{\frac{1}{2}}$ vs $V_{g}$ for pairs of $G(B)$ traces
separated by a period of $1$~hr at fixed $V_{g}$ showing the
increased MCF changes as the gate electric field shifts the device
further from its equilibrium acceptor charge configuration at
cooldown (i.e., $V_{g} = 0$).}\label{fig3}
\end{figure*}

\subsection{Temporal stability of the MCF}
\label{sec:temporal} The hysteresis in Fig.~2(d) indicates some
charge between the gate and 2DHG remains mobile even at $T~=~40$~mK.
The first question, by necessity, is the timescale over which the
charge configuration remains stable at fixed $V_{g}$ and minimum
$T$. This is essential to knowing whether $G(B)$ provides a
meaningful quasi-static magnetofingerprint of transport on the
timescale for obtaining a $G(B)$ trace, typically $\sim 20$~min, and
that $G(B)$ obeys known behaviors e.g., the Onsager-Casimir
symmetry relation $G(B) = G(-B)$.~\cite{OnsagerPR31, CasimirRMP45}
Note that the defined dot geometry is fixed, so MCF changes at fixed
$G(B~=~0)$ directly reflect changes in the dot's underlying disorder
potential.~\cite{SeePRL12, ScannellPRB12}

Figures~3(a/b) show how $G(B)$ evolves at fixed $V_{g}$, with traces
at times $t = 0$ (black), $8$ (blue) and $15$ hours (green) after
setting $V_{g} = -58.5$~mV and $-156$~mV to give (a) $G(B~=~0) \sim
1.8~G_{0}$ and (b) $G(B~=~0) \sim 2.5~G_{0}$, respectively. We
obtain almost identical MCF at $t = 0$, $8$ and $15$~hrs with $V_{g}
= -58.5$~mV (Fig.~3(a)), but MCF evolves markedly over time with
$V_{g} = -156$~mV (Fig.~3(b)), indicative of significant disorder
potential changes within the dot.~\cite{SeePRL12, ScannellPRB12} To
highlight this, Figures~3(c/d) present the conductance difference
$\Delta(B) = \delta G(B) - \langle \delta G(B) \rangle_{B}$ where
$\delta G(B) = G(B,t) - G(B,0)$ for $t = 8$ or $15$ hours. The
subtraction of $\langle \delta G(B) \rangle_{B}$ removes any
background d.c. conductance offset due to changes in trapped charge
outside the dot/QPCs and more than a coherence length away. This
offset is not always gradual; random `jumps' to higher/lower $G$
occur every few hours. One example occurred between the top and
middle traces in Fig.~3(a), removing the need for a vertical
offset for the blue trace. Thus, changes in $G(B)$ show up in
$\Delta(B)$ as fluctuations around $\Delta(B) = 0$.

Fig.~3(c) shows that $\Delta(B)$ remains close to $0$ for all $B$
with $V_{g} = -58.5$~mV, as expected for almost identical $G(B)$
traces. Conversely, substantial $\Delta(B)$ fluctuations emerge for
$V_{g} = -156$~mV (Fig.~3(d)), with the magnitude of the
fluctuations increasing with time. The marked changes to disorder
potential at more negative $V_{g}$ suggests the gate electric field
drives this underlying change in disorder potential. One possible
scenario is: cooling the device at $V_{g} = 0$ sets an initial
equilibrium state for trapped charge in the Si acceptor layer. The
equilibrium bandstructure in this finite thickness layer is such
that the occupied trap density in the growth direction $z$ is
inhomogeneous; this picture is consistent with observations for the
deep-trapping Si DX center in (100)-oriented AlGaAs/GaAs
heterostructures,~\cite{Pioro-LadrierePRB05, BuksSST94} albeit with
shallower traps here. A non-zero $V_{g}$ produces an electric field
along $z$ that drives charge migration between trap sites. If $kT <
E_{A}$ then the migration rate is slow and should increase with the
difference between $V_{g}$ and its value during cool-down (i.e.,
$V_{g} = 0$).

To test this scenario, we took pairs of $G(B)$ traces separated by
$1$~hr, starting at $V_{g} = 0$~V and using progressively more
negative $V_{g}$. To quantify MCF differences, i.e., the magnitude
of $\Delta (B)$ fluctuations, we plot the root-mean-square
conductance difference $\langle \Delta (B)^{2}
\rangle_{B}^{\frac{1}{2}}$ versus $V_{g}$ in Figure~3(e). The
quantity $\langle \Delta (B)^{2} \rangle_{B}^{\frac{1}{2}}$ is
similar to the correlation $F$ used previously,~\cite{SeePRL12,
ScannellPRB12, TaylorPRB97} and differs only by the absence of a
normalization coefficient based on random traces.~\cite{TaylorPRB97}
For completeness, corresponding $F$ values are shown in the
Appendix. Fig.~3(e) shows a clear increase in
$\langle \Delta (B)^{2} \rangle_{B}^{\frac{1}{2}}$ with more
negative $V_{g}$, consistent with gate field driven charge migration
in the Si acceptor layer. 

\begin{figure*}
\includegraphics[width=15cm]{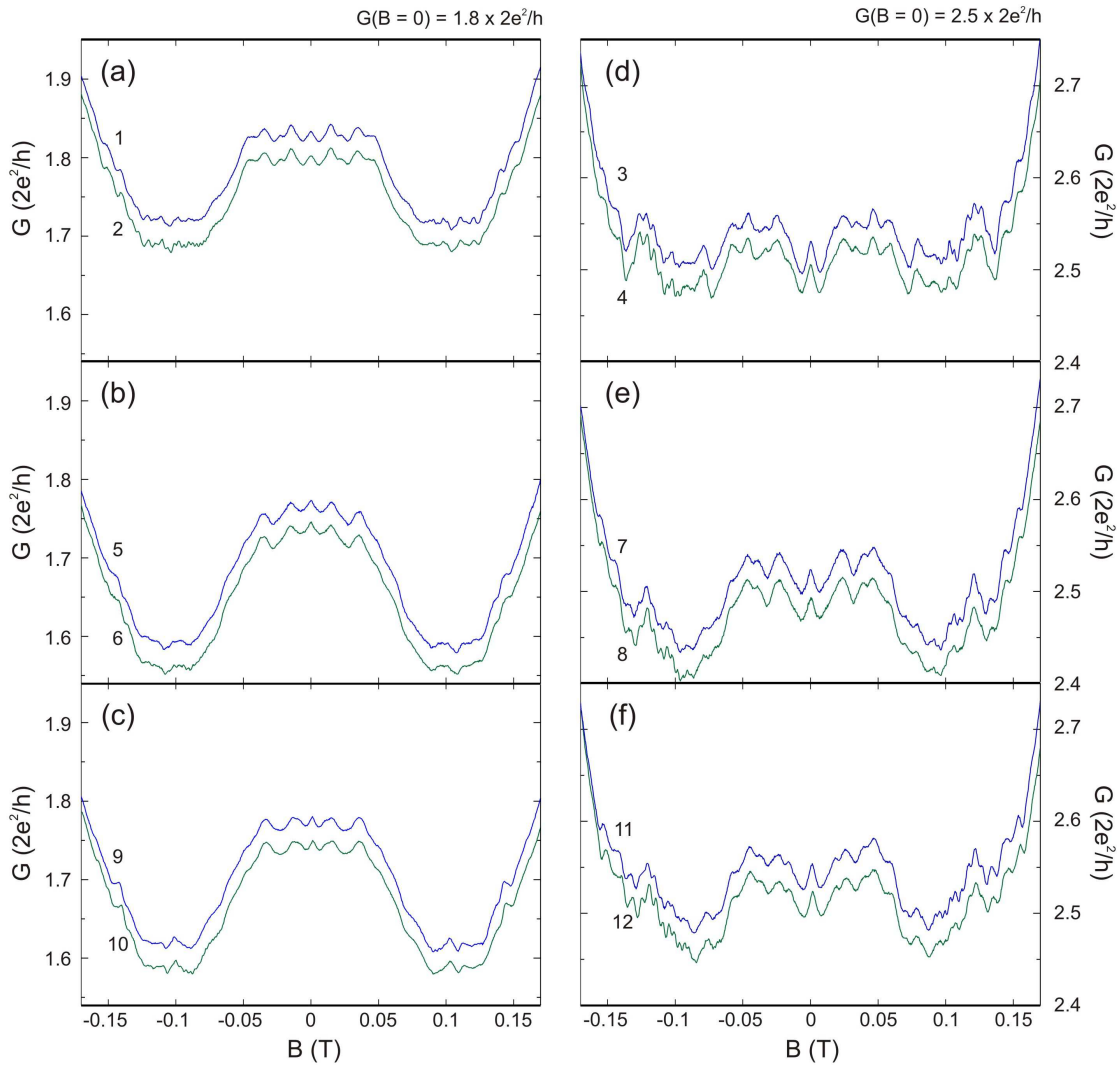}
\caption{(Color online) MCF traces obtained at (a-c) $G(B~=~0) = 1.8~G_{0}$ and
(d-f) $G(B~=~0) = 2.5~G_{0}$ with the numbers $1-12$ indicating the
order the traces were obtained in. The protocol is to: Set $G(B~=~0)
= 1.8~G_{0}$, obtain $G(B)$ [trace 1 in (a)], wait $1$~hr, obtain
$G(B)$ [trace 2 in (a)], set $G(B~=~0) = 2.5~G_{0}$, obtain $G(B)$
[trace 3 in (d)], wait $1$~hr, obtain $G(B)$ [trace 4 in (d)], set
$G(B~=~0) = 1.8~G_{0}$, obtain $G(B)$ [trace 5 in (b)], wait $1$~hr,
obtain $G(B)$ [trace 6 in (b)], and so on through the remaining $6$
traces. The second (green) trace in each panel is offset vertically
for clarity by $-0.03~G_{0}$.} \label{fig4}
\end{figure*}

\subsection{Evolution of MCF with gate modulation}
\label{sec:gateMCF} Figure~3(e) naturally leads to the question: How
robust is the MCF if $V_{g}$ is swept to some distant value and
returned to its original setting? In other words, to what extent
does charge migration in $z$ produce changes in charge distribution
in the $x-y$ plane, i.e., ionized dopant potential? This will be
linked to trap site population -- if traps are mostly filled, charge
may migrate in $z$ and ultimately return to the same initial trap
site under a cyclic variation in $V_{g}$; if they are mostly empty,
the probability of return will be low.

Figure~4 shows pairs of $G(B)$ traces obtained alternately at the
$G(B~=~0) = 1.8$ and $2.5~G_{0}$ plateaus. Two traces were taken at
each gate voltage to directly compare the temporal stability to
disorder induced by the gate potential over a similar time scale:
The numbers $1-12$ in Fig.~4 indicate the sequence in which traces
were obtained. The left (right) column in Fig.~4 should be
considered as a cyclic variation from $G = 1.8~G_{0} (2.5~G_{0})$ to
$G = 2.5~G_{0} (1.8~G_{0})$ and back; as such we do this experiment
twice, once starting at more positive $V_{g}$ and cycling to more
negative $V_{g}$ (left column), and once {\it vice versa} (right
column). If cyclic variation produced no change in the dot's
disorder potential, we would expect six identical $G(B)$ traces at
$G(B~=~0) = 1.8~G_{0}$, and a different set of six identical $G(B)$
traces at $G(B~=~0) = 2.5~G_{0}$; the difference in $G(B)$ between the
$G = 1.8 $ and $2.5~G_{0}$ plateaus is caused by the underlying 
difference in hole density. Inspection of Fig.~4 makes it clear the
disorder potential changes with cyclic variation in $V_{g}$.
Specifically, $G(B)$ varies more between panels than within each
panel, suggesting that cyclic variation in $V_{g}$ produces stronger
$G(B)$ changes than background temporal changes.

To confirm this, we again used $\langle \Delta (B)^{2}
\rangle_{B}^{\frac{1}{2}}$ to quantify the $G(B)$ changes: Pairs of
traces in each panel give an average $\langle \Delta (B)^{2}
\rangle_{B}^{\frac{1}{2}}$ of $1.1~\times~10^{-3}~G_{0}$
($2.6~\times~10^{-3}~G_{0})$ for the left (right) column of
Fig.~4. In contrast, if we compare the second traces in a/b, b/c
and a/c (or d/e, e/f and d/f) we get an average $\langle \Delta
(B)^{2} \rangle_{B}^{\frac{1}{2}}$ of $10.3~\times~10^{-3}~G_{0}$
($9.7~\times~10^{-3}~G_{0})$. The $\langle \Delta (B)^{2}
\rangle_{B}^{\frac{1}{2}}$ values relating to cyclic variation are
$5-10\times$ larger than those relating to temporal changes: this
confirms that significant changes in dot disorder potential result
from cyclic variation in $V_{g}$. The fact that the MCF continues 
to evolve through the 6 cycles also suggests that the trap site 
population is low in this sample. This is not surprising, as the 
primary purpose of the dopant charge is to populate the 2DHG and quantum dot.

\begin{figure*}
\includegraphics[width=15cm]{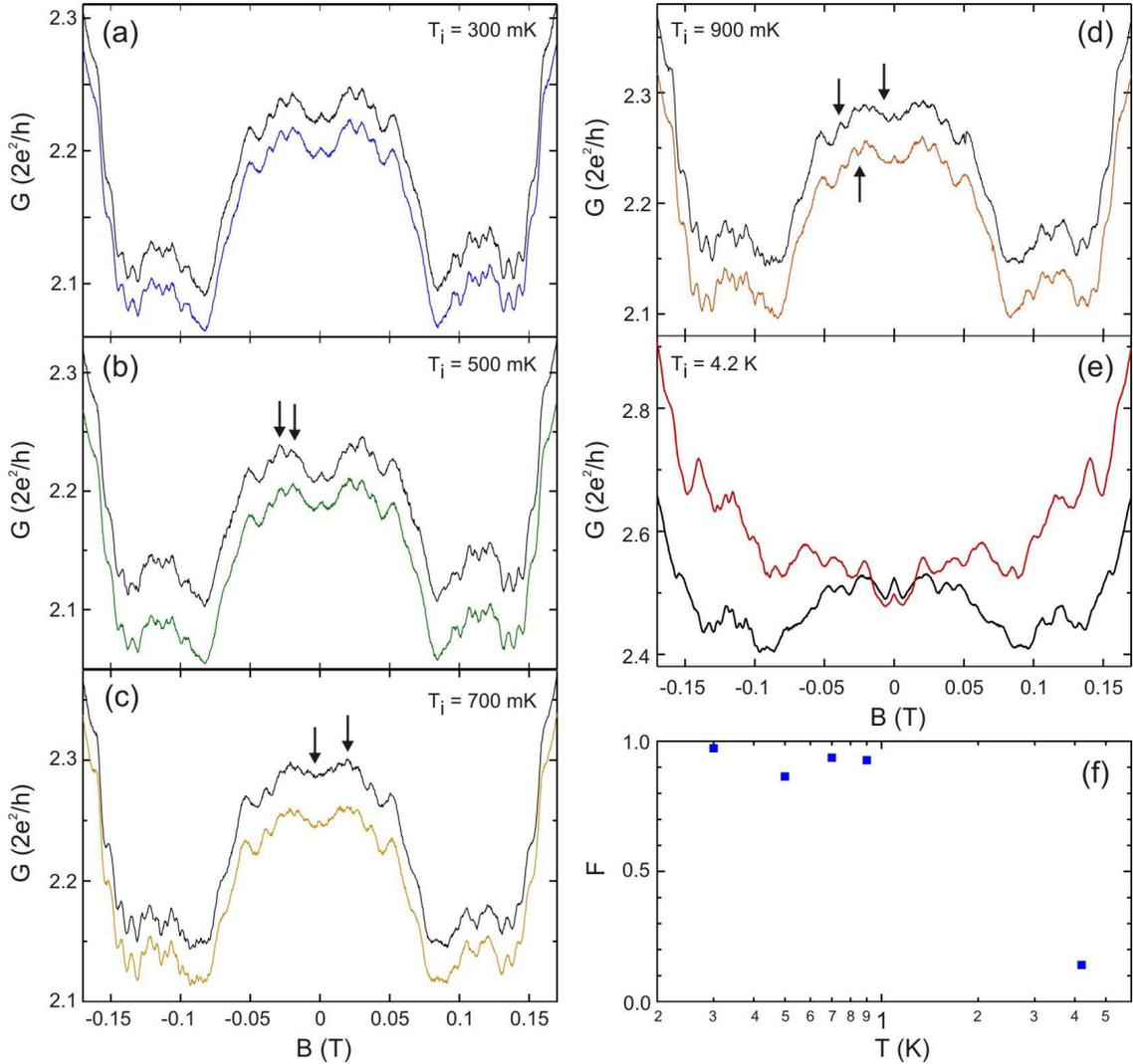}
\caption{(Color online) MCF traces before and after (upper and lower trace in each
frame, respectively) cycling to intermediate temperatures ($T_{i}$)
of (a) $300$~mK (b) $500$~mK (c) $700$~mK (d) $900$~mK and (e)
$4.2$~K. Traces in (a - d) were obtained with $V_{g} = 0$, and
$T_{i}$ was held for $30$~mins. For (e), $V_{g} = -0.156$ and
$-0.270$~V for the initial and final traces, respectively, and the
temperature variation was more complex, as discussed in the text.
The lower traces in (a), (b) and (c) have a vertical offset of
$-0.01~G_{0}$, $-0.04~G_{0}$ and $-0.03~G_{0}$, respectively. The
differences in $G(B~=~0)$ between traces in each experiment were (a)
$-0.017~G_{0}$, (b) $0.015~G_{0}$, (c) $0.007~G_{0}$, (d)
$-0.037~G_{0}$ and (e) $0.03~G_{0}$. Cycling to $300$~mK produces no
change in the MCF, but changes emerge upon cycling to $\geqslant
500$~mK. (f) The correlation $F$ vs $T_{i}$ for pairs of $G(B)$
traces in (a - e).}\label{fig5}
\end{figure*}

\subsection{Evolution of MCF with thermal cycling}
\label{sec:temperature} The previous two sections showed that at $T
= 40$~mK, charge migration occurs between some proportion of Si
acceptors in the AlGaAs layer under the influence of an electric
field, but spontaneous re-organization does not occur. This suggests
a metastable configuration: i.e., the thermal energy $kT$ at $40$~mK
is slightly lower than the activation energy $E_{A}$ for the Si
acceptors. We thus attempted to determine $E_{A}$ via temperature
studies of $G(B)$. The probability of trap deoccupation depends on
$kT$ relative to $E_{A}$, thus elevating $T$ should lead to more
rapid changes in disorder potential and thereby in $G(B)$, reflected
by increased $\langle \Delta (B)^{2} \rangle_{B}^{\frac{1}{2}}$.
Scannell {\it et al.}~\cite{ScannellPRB12} recently performed such a
study for GaAs and InGaAs electron quantum dots using the following
methodology: Set and hold a particular $V_{g}$, obtain $G(B)$ at
base temperature $T_{0} = 300$~mK, heat to $T_{1} >> 300$~mK for
some time $t$, return to $T_{0}$ and remeasure $G(B)$, heat to
$T_{2} > T_{1}$ for $t$, return to $T_{0}$ and remeasure $G(B)$,
repeat $i$ times until $T_{i} = 300$~K. The $G(B)$ at each iteration
$i$ is then compared to the initial $G(B)$ trace, with the
correlation $F = 1 - \langle \Delta (B)^{2}
\rangle_{B}^{\frac{1}{2}}/N$, where $N$ is a normalization factor,
plotted against $T_{i}$ to identify where $kT_{i}$ becomes
sufficient to cause major changes to $G(B)$.~\cite{ScannellPRB12}

Repeating the study by Scannell {\it et al.} for a hole quantum dot,
as we do here, faces two serious challenges. The higher effective
mass for holes results in a significantly reduced phase coherence
time relative to electrons.~\cite{FanielPRB07} As a result $G(B)$
must be measured at lower $T$ to obtain MCF; the MCF of our device
is quenched at $T \gtrsim 100$~mK. This means our measurements
require a dilution refrigerator, where the available $T_{i}$ range
is limited to $T_{i} < 900$~mK, $T_{i} = 4.2$~K and $T_{i} = 300$~K.
In contrast, the $^{3}$He system used in Ref.~\cite{ScannellPRB12} allows
continuous, precise variation of $T_{i}$ over the range $300$~mK~$<
T_{i} < 200$~K. The second challenge is that $E_{A}$ is much smaller
for our system. This means that even if we had good control over
$T_{i}$ above $900$~mK, much of this range is of limited use as we
should reach the point where $\langle \Delta (B)^{2}
\rangle_{B}^{\frac{1}{2}}$ saturates at its maximum value at
relatively low $T_{i}$. Nonetheless, we attempt the study within
these limitations to confirm at least that $E_{A}$ for Si acceptors
in (311)A AlGaAs/GaAs heterostructures is much smaller than for Si
donors in (100) AlGaAs/GaAs heterostructures.

Figure~5 shows $G(B)$ traces obtained before and after cycling to
$T_{i} = 0.3, 0.5, 0.7, 0.9$ and $4.2$~K. We first make one
important note regarding interpretation relative
to Ref.~\cite{ScannellPRB12}; here we compare the $G(B)$ traces obtained
immediately before/after cycling to $T_{i}$, rather than comparing
$G(B)$ after to a common, initial `$T_{0}$' $G(B)$ trace. This is to
more fairly deal with the continuous temporal evolution/relaxation
in $G(B)$; i.e., we do not want our analysis to include the temporal
changes in $G(B)$ observed in Fig.~3. The MCF in Fig.~5(a) shows
little change after thermal cycling to $T_{i} = 300$~mK. For this
pair of traces we obtain a correlation $F = 0.97$. As we increase
$T_{i}$ in steps of $200$~mK in Fig.~5(b-d), the differences
between MCF traces before and after cycling to $T_{i}$ begin to
increase, as indicated by the arrows. Correspondingly, there is a
gradual fall-off in $F$ towards $\sim 0.93$, evident in Fig.~5(f)
where we plot $F$ versus $T_{i}$. There is a little scatter in $F$
for $T_{i} < 900$~mK, but this is also apparent
in Ref.~\cite{ScannellPRB12}, and often results from small, slow
background variations running through the data. Ultimately, upon
reaching $T_{i} = 900$~mK, $F$ has not dropped off markedly.

While the data in Fig.~5(a-d) could be obtained at fixed $V_{g}$,
with $T$ controllably increased to $T_{i}$, held there for $30$~min
and then returned to $T = 40$~mK for $G(B)$ measurement, this is not
possible for the next available $T_{i} = 4.2$~K. More effort is
needed to reach $4.2$~K with a dilution refrigerator, i.e., remove
of $^{3}$He/$^{4}$He mixture, add exchange gas, equilibrate with
$4.2$~K bath, pump out exchange gas, recondense mixture and restart
circulation; this process takes many hours. Additionally, the
resulting dopant reorganization inevitably means that $G \sim
2.5~G_{0}$ does not occur at the same $V_{g}$. Indeed, the initial
and final $G(B)$ traces in Fig.~5(e) were obtained at $V_{g} =
-0.156$ and $-0.270$~V, respectively. They are markedly different,
and the corresponding $F = 0.14$.

Fig.~5(f) shows that the $T_{i}$ where the MCF decorrelates
rapidly lies somewhere in the range $0.9 < T_{i} < 4.2$~K. The rapid
$F$ drop-off for an AlGaAs/GaAs electron dot in Ref.~\cite{ScannellPRB12}
occurs at $T_{i} = 165$~K. If we estimate that $T_{i} \sim 2$~K for
$F = 0.5$ in Fig.~5(f), then we would expect the trap depth for
our Si acceptors to be $1.2\%$ of that for Si donors. Assuming the
Si donors are DX centers with depth $\sim
250$~meV,~\cite{MooneyJAP90} then we can estimate the shallow Si
acceptor trap depth for our dots to be $E_{A} = 3$~meV. The isolated
Si acceptor energy in GaAs is $34.8$~meV~\cite{KirkmanJPC78} and
$\sim 60$~meV in AlGaAs~\cite{GalbiatiSST96}; our trap $E_{A}$ is
over an order of magnitude smaller. This value of $E_{A} = 3$~meV
should be considered a lower bound estimate as the difficulties
involved in obtaining the $T_{i} = 4.2$~K measurement could have
reduced the measured $F$. However, even if the decorrelation
temperature were as high as $T = 20$~K, the corresponding $E_{A} =
30$~meV would still be $\sim 2 \times$ smaller than the isolated
Si-acceptor level in AlGaAs. One possible source of this small trap
is the Si-X acceptor defect,~\cite{MurrayJAP89} which appears at
very high Si doping densities. As Ashwin {\it et al.} point out,
Si-X would need a small activation energy to be electrically active
amongst a high density of isolated Si acceptors.~\cite{AshwinJAP94}

\section{Conclusions}
We have used the sensitivity of the magnetoconductance fluctuations
in an open quantum dot to small-angle scattering from remote ionized
impurities~\cite{SeePRL12, ScannellPRB12} to study the stability of
Si acceptors in (311)A AlGaAs/GaAs heterostructures. With $T =
40$~mK and $V_{g} \sim 0$, the MCF remain reproducible for a period
$\sim 15$~hrs, demonstrating that under the simplest quasi-static
conditions, a hole quantum dot can show stable, meaningful
electrical characteristics and MCF, consistent with previous
studies.~\cite{FanielPRB07} By contrast, the temporal stability of
the MCF is significantly reduced as $V_{g}$ is made more negative.
Similarly, MCF changes are observed when cycling $V_{g}$ away from
some initial value, and returning to a value that gives the same
$G(B~=~0)$. We attribute this to migration of charge between trap
sites in the Si acceptor layer due to the electric field induced
between the gate and 2DHG by $V_{g}$. These results suggest a
metastable state for Si acceptors in AlGaAs at $T = 40$~mK, with the
activation energy $E_{A}$ of acceptors larger than the thermal energy
$kT$. To determine $E_{A}$, we followed Scannell {\it et al.} and
looked at how periods at higher temperatures $T_{i}$ changed the MCF
at fixed $V_{g}$. This experiment faced many difficulties, but we
found that a $T_{i} = 4.2$~K is sufficient to produce complete
decorrelation of the MCF. By comparing with the known trap energy
for Si DX centers and the $T_{i}$ required to produce decorrelation
in electron devices, we estimate that the shallow acceptor traps in
our heterostructures have an activation energy $E_{A} \sim 3$~meV,
consistent with our earlier studies of device
hysteresis.~\cite{BurkePRB12} This work demonstrates the capacity
for the MCF in quantum dots to be used as a tool for detecting
temporal changes in ionized dopant configuration and charge trap
occupation in semiconductor heterostructures. One structure
of immediate interest is C-doped $p$-type AlGaAs/GaAs
heterostructures, since devices fabricated from this material also exhibit
gate instability at low temperature.~\cite{CsontosAPL10} Repeating the
experiments conducted here for this material could help to determine
whether charge migration between C acceptors plays a role in this
instability, and determine the C acceptor activation energy in AlGaAs.

\begin{acknowledgements}
This work was funded by Australian Research Council Grants 
DP0110103802, FT0990285, DP120101859, DP120102888 and by 
the Australian Government under the Australia-India Strategic 
Research Fund. DR and ADW acknowledge support
from DFG SPP1285 and BMBF QuaHL-Rep 16BQ1035. This work was
performed in part using the NSW node of the Australian National
Fabrication Facility (ANFF). We thank L. Eaves for helpful
discussions on donors and acceptors in AlGaAs/GaAs heterostructures.
\end{acknowledgements}

\appendix*
\section{Correlation data for Figure 3(e)}
Figure~6 shows the correlation $F$ versus gate voltage $V_{g}$
corresponding to the data presented in Figure~3(e).
The correlation $F$ is obtained by applying a cross-correlation
analysis to a pair of traces $G_{1}(B)$ and $G_{2}(B)$:

\begin{eqnarray}
F &=& \sqrt{1-\frac{\left\langle\left[G_1(B)-G_2(B)\right]\right\rangle_B}{N}}~~~\textrm{where} \\
N &=& \left\langle\left[G_x(B)-G_y(B)\right]\right\rangle_B.
\end{eqnarray}

\noindent The symbol $\langle~\rangle_B$ represents an average over
the $14001$ data points spanning $-B_{c} < B < B_{c}$, where $B_{c}
= 0.07$~T is the field at which the cyclotron diameter matches the
billiard width. The normalization constant $N$ is calculated by
averaging the correlations of $100$ pairs of randomly generated
$G(B)$ traces with the same average, maximum and minimum $G$ as the
two analysed traces $G_{1}(B)$ and $G_{2}(B)$. Adopting this
normalization, the correlation scale varies between 1 for
mathematically identical traces to 0 for complete decorrelation.
Further details of this method can be found in
Refs.~\cite{TaylorPRB97, SeePRL12, ScannellPRB12}.

\begin{figure}
\includegraphics[width=8cm]{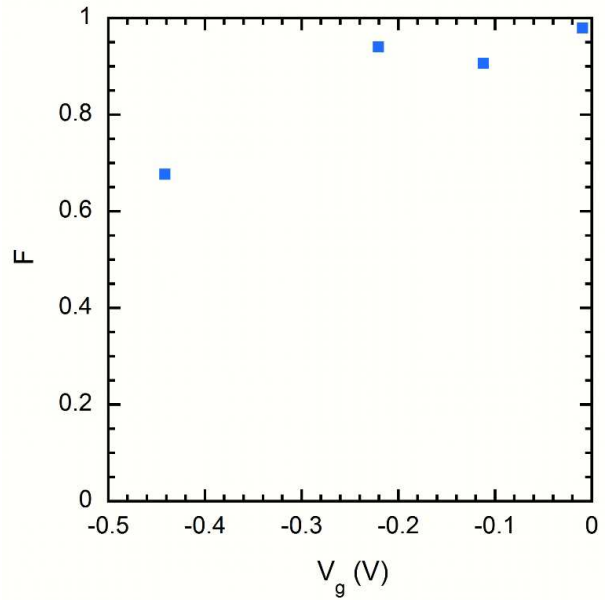}
\caption{Correlation $F$ vs gate voltage $V_{g}$ corresponding to
the data presented in Fig.~3(e) of the main text.}
\end{figure}


%

\end{document}